\documentclass[11pt,a4paper]{article}
\usepackage{epsfig}

\newlength{\dinwidth}
\newlength{\dinmargin}
\def\be{\begin{equation}}
\def\ee{\end{equation}}
\def\bear{\begin{eqnarray}}
\def\eear{\end{eqnarray}}
\def\bfi{\begin{figure}}
\def\efi{\end{figure}}
\def\btab{\begin{table}}
\def\etab{\end{table}}
\def\bc{\begin{center}}
\def\ec{\end{center}}
\def\bi{\begin{itemize}}
\def\ei{\end{itemize}}
\def\nn{\nonumber}

\def\text{\textstyle}

\def\mathswitch#1{\relax\ifmmode#1\else$#1$\fi}

\newcommand{\tosv}
{{\scriptscriptstyle \to}}

\newcommand{\qbar}{{\bar q}}
\newcommand{\gl}{\tilde{g}}
\newcommand{\sq}{\tilde{q}}
\newcommand{\sqb}{\bar{\tilde{q}}}
\newcommand{\as}{\alpha_{\mathrm{s}}}
\newcommand{\si}{\mathswitch \sigma}
\newcommand{\sih}{\mathswitch \hat \sigma}

\newcommand{\nf}{\mathswitch {n_{\mathrm{f}}}}

\newcommand{\CA}{\mathswitch {C_{\mathrm{A}}}}
\newcommand{\CF}{\mathswitch {C_{\mathrm{F}}}}
\newcommand{\shat}{{\hat s}}

\newcommand{\msq}{\mathswitch m_{\sq}}
\newcommand{\mgl}{\mathswitch m_{\gl}}
\newcommand{\rhohat}{\mathswitch \hat{\rho}}

\newcommand{\NLL}{\mathrm{NLL}}

\newcommand{\NLO}{\mathrm{NLO}}
\newcommand{\etal}{{\it et al.}}

\newcommand{\onebf}{{\bf 1}}
\newcommand{\eigbf}{{\bf 8}}
\newcommand{\eigbfa}{{\bf 8_{\mathrm \bf A}}}
\newcommand{\eigbfs}{{\bf 8_{\mathrm \bf S}}}
\newcommand{\tptbf}{{\bf 10}\oplus {\bf \overline{10}}}
\newcommand{\twsbf}{{\bf 27}}

\newcommand{\ord}{{\cal O}}

\newcommand{\GeV}{\unskip\,\mathrm{GeV}}

\newcommand{\TeV}{\unskip\,\mathrm{TeV}}
\newcommand{\fba}{\unskip\,\mathrm{fb}}

\newcommand{\BA}{\Lambda}
\newcommand{\C}{\Omega}
\newcommand{\mre}{\mathrm{Re}\,}
\newcommand{\Sgl}{\bar{S}}
\newcommand{\Tgl}{\bar{T}}
\newcommand{\Ugl}{\bar{U}}
\newcommand{\A}{\Sgl}

\begin{document}

\thispagestyle{empty}
\def\thefootnote{\fnsymbol{footnote}}
\setcounter{footnote}{1}
\null
\date
\strut\hfill   DESY 08--096\\
\strut\hfill arXiv:0807.2405
\vskip 0cm
\vfill
\begin{center}

{\Large \bf Threshold resummation for squark-antisquark and gluino-pair 
production at the LHC}

\bigskip
\bigskip
\bigskip
\bigskip

{\large \sc
A.~Kulesza$^a$ and L.~Motyka$^{b,c}$}
~\footnote{% 
{anna.kulesza@desy.de},
{leszek.motyka@desy.de}}

\bigskip
\bigskip

\begin{it}

$^a$Deutsches Elektronen-Synchrotron DESY, Notkestrasse 85,\\
D--22607 Hamburg, Germany \\
\bigskip

$^b$II Institute for  Theoretical Physics, 
University of Hamburg,\\ Luruper Chaussee 149, D-22761, Germany \\
\bigskip

$^c$Institute of Physics, Jagellonian University, Reymonta 4,\\ 30-059
Krak\'{o}w, Poland

\end{it}

\bigskip
\bigskip

\centerline{July 15, 2008}

\bigskip
\bigskip
\bigskip

\end{center}

\vfill
\bc
{\bf Abstract}
\ec

We study the effect of soft gluon emission in the hadroproduction of
squark-antisquark and gluino-gluino pairs at the next-to-leading
logarithmic (NLL) accuracy within the framework of the minimal
supersymmetric model. 
The one-loop soft anomalous dimension matrices controlling the colour
evolution of the underlying hard-scattering processes are calculated. 
We present the resummed total cross sections and show numerical results 
for proton-proton collisions at 14 $\TeV$. The size of the NLL contribution 
to the cross section and the reduction of the scale dependence 
of the theoretical predictions due to including soft gluon
effects are discussed.

\par
\vskip 1cm
\noindent
\par
\null
\setcounter{page}{0}
\clearpage
\def\thefootnote{\arabic{footnote}}
\setcounter{footnote}{0}

\newpage

\section{Introduction}
The search for new physics phenomena at the TeV scale begins 
at the Large Hadron Collider (LHC) this year. 
Among the proposed models of new physics,
extentions of the Standard Model (SM) involving supersymmetry (SUSY) are 
one of the best motivated. Over the years much attention has been focused 
on the Minimal Supersymmetric Standard Model (MSSM)~\cite{mssm}, 
characterized by the minimal content of supersymmetric particles and 
$R$-parity conservation. In the MSSM the SUSY partners of the SM 
particles are always produced in pairs.

The dominant production processes of sparticles at the LHC 
are those involving pairs of coloured 
particles, i.e.\ squarks  
and gluinos, in the final state~\cite{susyrev}.
The discovery
of squarks and gluinos should be possible for 
masses of up to 2~TeV \cite{tdrs}.
Depending on the outcome of the experimental searches, predictions for 
the total rates for these production processes will either
help to determine the masses of the sparticles~\cite{Baer} or to draw 
exclusion limits for the mass parameters.
It is therefore important to know the LHC cross sections for
production of squarks and gluinos with high
theoretical accuracy. In particular, control over effects of
higher-order radiative corrections leads to a stabilization of the
theoretical predictions w.r.t.\ variation of the renormalisation 
and factorisation scales and, consequently, to a smaller theoretical 
error.

The leading-order (LO) total cross sections and the corresponding next-to-leading order (NLO)
SUSY-QCD corrections are known for all hadroproduction processes of pairs 
of squarks and gluinos~\cite{DEQ,BHSZ1,BHSZ2}.
They have been found to be positive and large, especially for the
gluino-pair production. As pointed out in~\cite{BHSZ2}, an important
part of the contributions to the hadronic cross sections comes from
the energy region near the partonic production threshold. 
The threshold region is reached when the square of the partonic 
center-of-mass (c.o.m.) energy, $\shat$, approaches $4\,m^2$, 
where $m$ is the average particle mass in the produced pair.
The velocity of the produced heavy 
particles in the partonic c.o.m.\  system 
$\beta \equiv \sqrt{1- 4 m^2/ \shat}$ is then small, $\beta \ll 1$. 
In this region, two types of corrections dominate: 
Coulomb corrections, due to exchange of gluons between
slowly moving massive particles, and soft gluon corrections, due to
emission of low energy gluons off the coloured initial and
final states.  
The large size of the soft gluon emission contributions can be traced
down, for the perturbative $n$-th order correction,  to the logarithmic terms of the form $\as^n \log^{k}(\beta^2)$ where
$k=2n, \dots, 0$.
The effects of
the soft gluon emission can be taken into account to all orders in
perturbation theory by performing 
resummation of the threshold logarithms. Resummed predictions are particularly 
important for processes with large masses in the final states, since then 
the bulk of the production comes from the threshold region. This is exactly the
case for the production of the sparticles  which are expected to
be heavier than the SM particles. Additionally, if the partonic
subprocesses involve gluons in the initial state, the soft gluon effects, and
thus the impact of resummation, are
expected to be significant due to the high colour charge of the gluons.  

In this letter, we report on the calculation of  
threshold-resummed cross sections for 
hadroproduction of gluino-gluino ($\gl\gl$)  and squark-antisquark 
($\sq \sqb$) pairs.  Among the pair-production processes of coloured
sparticles at the LHC, the $\gl\gl$ production receives the 
largest NLO SUSY-QCD
correction~\cite{BHSZ2}, and the NLO $K$-factor, 
$K_{\NLO}$, can reach~2 for gluino mass $m_{\gl} = 1$~TeV. The corrections 
to the $\sq \sqb$ total cross section can be also sizeable $K_{\NLO} \simeq 1.3$ 
for the squark mass, $m_{\sq}=1$~TeV, and are the second largest in a
certain range of mass parameters. The hadronic production of $\gl\gl$ and 
$\sq \sqb$ pairs are scattering processes with a 
non-trivial colour flow structure. At the level of next-to-leading logarithms (NLL), resummation 
requires including contributions from soft gluons emitted at wide angles. Such
emission is sensitive to the colour flow of the underlying hard scattering and
the evolution of the colour exchange is governed by the 
one-loop 
soft anomalous dimension matrix~\cite{BS,KS,KOSthr,KOScol,BCMNgen}. 
These matrices have been calculated for heavy-quark and 
dijet production~\cite{KS,BCMNtop,BS,KOScol} and the general 
results for any $2\to n\,$ QCD process with massless particles
in the final state have been derived at one-~\cite{BCMNgen}, 
and two-loops~\cite{MDS}. Recently, the two-loop anomalous dimension for the pair 
production of heavy quarks has been also determined in the threshold limit~\cite{MU}.
Here we present the explicit form of the soft anomalous dimension matrices
for partonic subprocesses contributing to $\gl\gl$ hadroproduction 
and apply the resummation formalism at the NLL level to evaluate the
correction due to soft gluon emission for  the production of 
$\gl\gl$ and $\sq\sqb$ pairs at the LHC.

\section{Soft gluon resummation}
In this letter we consider the hadronic production of 
gluino-pair and squark-antisquark final states, \mbox{$h_1 h_2 \to \gl \gl$}
and \mbox{$h_1 h_2 \to \sq \sqb$}, where we sum over the left- and right-
chiralities of squarks and over the spins of gluinos. At LO, these two
processes receive contributions from the following partonic channels: 
$$
q \qbar \to \gl \gl, \qquad\qbar q
  \to \gl \gl,\qquad gg \to \gl \gl
$$ 
and
$$q \qbar \to \sq \sqb,\qquad \qbar q
  \to \sq \sqb, \qquad gg \to \sq \sqb,
$$ 
respectively. 
The resummation of the soft gluon corrections is carried out in
Mellin-$N$ space  in the variable \mbox{$\rho= 4 m^2/S$} with $S$ being the 
square of the hadronic c.o.m energy and $m$ the mass of the final-state particle.
In the Mellin space, the moments of the partonic cross section $ij \to kl$ are
given by 
\be
\sih_{ij \tosv kl,N}(m^2, \mu^2_F,\mu^2_R) \;\equiv\; 
\int_0^1 d \rhohat \;
\rhohat^{N-1} \; 
\sih_{ij \tosv kl} (\rhohat, m^2,\mu^2_F,\mu^2_R)\,,
\ee
with $\rhohat= 4 m^2/\shat$.

The evolution of colour exchange 
in non-collinear soft gluon emission, which has to be taken into account at
the NLL accuracy,  
is governed by the soft anomalous dimension
matrix $\Gamma^{ij \tosv kl}$~\cite{BS,KS,KOSthr,KOScol,BCMNgen}.
In an orthogonal basis in the colour space for which the matrix  
$\Gamma^{ij \tosv  kl}$ is diagonal, the NLL resummed cross section in the
$N$-space has the form~\cite{KS,BCMNtop}
\be
\label{eq:res:fact}
\sih^{{\rm (res)}}_{ij\tosv kl,N} \; = \;
\sum_{I} \sih^{(0)}_{ij\tosv  kl,I,N}\; \widetilde{C}_{ij\tosv kl,I}\;
\Delta^i_{N+1} \; \Delta^j_{N+1} \; 
\Delta^{\rm (int)}_{ij\tosv kl,I,N+1}\, ,
\ee
where we suppress explicit dependence on the scales. 
The index $I$ in Eq.~(\ref{eq:res:fact}) distinguishes between contributions from
different colour channels. The colour-channel-dependent contributions
to the LO partonic cross sections in Mellin-moment space are
denoted by $\sih^{(0)}_{ij\tosv kl,I,N}$ and will be presented 
elsewhere~\cite{inpreparation}. The radiative factors
$\Delta^i_{N}$ describe the effect of the soft gluon radiation
collinear to the initial state partons and are universal. Large-angle
soft gluon emission is accounted for by the factors $\Delta^{\rm
    (int)}_{ ij\tosv kl,I,N}$ which depend on 
the partonic process under consideration and the colour configuration of
the participating particles. The functions $\widetilde{C}_{ij\tosv kl,I}$ are
in general obtained by comparison of the resummed formulas with the
fixed-order expressions. They consist of Coulomb corrections and $N$-independent terms (often referred to in the literature under the common 
symbol $C_{ij\tosv  kl}$), containing hard contributions from virtual
corrections. For the most complete treatment of the threshold effects, 
the Coulomb corrections should be resummed, see e.g.~\cite{CMNTCoul,BCMNtop}. 
The concurrent treatment of the Coulomb and soft gluon corrections is, 
however, beyond the scope of the work reported on here.

The expressions for the radiative factors in the $\overline{\mathrm{MS}}$ 
factorisation scheme read~(see e.g.~\cite{BCMNtop})              
\bear
\ln \Delta^{i}_N  & = & \int_0^1  dz \,\frac{z^{N-1}-1}{1-z} 
\,
\int_{\mu_{F}^2}^{ 4m^2(1-z)^2} \frac{dq^2}{q^2} A_i(\as(q^2))\, , \nn \\
\ln \Delta^{\rm (int)}_{ij\tosv kl,I,N} &  = &
 \int_0^1  dz \, \frac{z^{N-1}-1}{1-z} \, 
D_{ ij\tosv kl,I}(\as(4m^2(1-z)^2)) \, . 
\eear
 The coefficients $\;{\mathcal F}= A_i,\,D_{ ij\tosv kl,I}\;$ are 
power series in the coupling constant $\,\as\,$,
%\mbox{
$\;{\mathcal F}= (\frac{\as}{\pi}){\mathcal F}^{(1)}+(\frac{\as}{\pi})^2 
{\mathcal F}^{(2)}+ \dots$.
The universal LL and NLL coefficients $A_i^{(1)}$, $A_i^{(2)}$ are
well known~\cite{KT,CET} and given by
\bear
A_i^{(1)}\,&=&\, C_i\,, \nn \\
A_i^{(2)}\,&=&\, \frac{1}{2} \; C_i \left(\left( 
\frac{67}{18} - \frac{\pi^2}{6}\right) \CA - \frac{5}{9} \nf \right)
\eear  
with  \mbox{$C_g=\CA=3$,} and \mbox{$C_q=\CF=4/3$}.
The customary NLL expansions of the radiative factors can be 
found in~\cite{BCMNtop}. Since after the expansion of the exponentials 
the terms constant in $N$ contained in
$C_{ij \tosv kl}$ generate contributions of the next-to-next-to-leading
logarithmic (NNLL) order, and we do not include
Coulomb corrections here, we keep ${\widetilde{C}}_{ij\tosv kl,I}=1$ 
for the purpose of the calculations.

The NLL coefficients  $D^{(1)}_{ij\tosv \sq \sqb,\,I}$ 
are the same as in the heavy-quark production and were calculated
in~\cite{KS,BCMNtop}
\bear
D^{(1)}_{q\qbar \tosv \sq \sqb,1} 
&\,=\,&  D^{(1)}_{ gg \tosv \sq
   \sqb,1} \,=\, 0 \,, \nn \\
D^{(1)}_{q \qbar \tosv \sq \sqb,2} &\,=\,&  
D^{(1)}_{ gg \tosv \sq \sqb,2} \,=\, -\CA,
\eear
with the index $I\!=\!1,2$ corresponding to the singlet and octet~$\{\onebf,\eigbf\}$ exchange in the $s$-channel. 
For the $\gl\gl$ production we first
study the soft anomalous dimension matrices $\Gamma^{ij \to \gl \gl}$ and 
then derive the coefficients $D^{(1)}_{ij\to\gl\gl,I}$. Apart from obvious
colour-dependent modifications, the matrices $\Gamma^{ij \to \gl \gl}$
are defined in the same way as the soft anomalous dimension matrices for the
heavy-quark production process in~\cite{KS}.
For the purpose of this calculation we consider the partonic processes
$$
q(p_1,\alpha_1) \,\qbar(p_2,\alpha_2) \to \gl(p_3,a_3)\,\gl(p_4,a_4) 
$$
and
$$
g(p_1,a_1) \,g(p_2,a_2) \to \gl(p_3,a_3)\,\gl(p_4,a_4)\,,
$$
where $p_i$ are the particle four-momenta and $\alpha_i$ and $a_i$ are
the colour indices in the fundamental and adjoint representation of
SU(3), respectively.

Three independent tensors are forming the
basis in the space of colour exchanges for the $q\bar q \to \gl\gl$ process.
We choose an orthogonal $s$-channel basis \mbox{$\{c^q _I\}, \;I=1,2,3$:}
\be
c^q _{1} = \delta^{\alpha_1 \alpha_2} \, \delta^{a_3 a_4}, 
\quad
c^q _{2} = T^b _{\alpha_2 \alpha_1} d^{b a_3 a_4},
\quad
c^q _{3} = i T^b _{\alpha_2 \alpha_1} f^{b a_3 a_4},
\ee
where $T^{b}$ matrices are the SU(3) generators, and the tensors 
correspond to  $\onebf$, $\eigbfs$ and $\eigbfa$ representations,
respectively. 
For the  $gg$ channel there are eight independent 
colour tensors. Following~\cite{KOScol} we choose an orthogonal basis
\mbox{$\{c^g_I\}$} (\mbox{$I=1,2,\ldots,8$})
consisting of five  tensors $c^g _1$, $c^g _2$, $c^g _3$, $c^g _4$,
$c^g _5$ corresponding to the $\onebf$, $\eigbfs$, $\eigbfa$, $\tptbf$, $\twsbf$  
representations in the $s$-channel, and three additional tensors
$c^g _6$, $c^g _7$, $c^g_8$. The base tensors are
\bear
c^g _1 &=& {1\over 8}\delta^{a_1 a_2} \delta^{a_3 a_4},\nn \\ \nn
c^g _2 &=& {3\over 5}d^{a_1 a_2 b} d^{b a_3 a_4},\\ \nn
c^g _3 &=& {1\over 3}f^{a_1 a_2 b} f^{b a_3 a_4},\\ \nn
c^g _4 &=& {1\over 2} \left( \delta^{a_1 a_3}\delta^{a_2 a_4}
                     -  \delta^{a_1 a_4}\delta^{a_2 a_3} \right)
                       -{1\over 3} f^{a_1 a_2 b} f^{b a_3 a_4}, \\ \nn
c^g _5 &=&  {1\over 2} \left( \delta^{a_1 a_3} \delta^{a_2 a_4} 
       + \delta^{a_1 a_4} \delta^{a_2 a_3} \right)
        -{1\over 8}\delta^{a_1 a_2} \delta^{a_3 a_4}
        -{3\over 5}d^{a_1 a_2 b} d^{b a_3 a_4}, \\ \nn
c^g _6 &=& {i\over 4} \left(
f^{a_1 a_2 b} d^{b a_3 a_4} + d^{a_1 a_2 b} f^{b a_3 a_4} \right), \\\nn
c^g _7 &=& {i\over 4} \left(
f^{a_1 a_2 b} d^{b a_3 a_4} - d^{a_1 a_2 b} f^{b a_3 a_4} \right), \\ 
c^g _8  &=& {i\over 4} \left(
d^{a_1 a_3 b} f^{b a_2 a_4} + f^{a_1 a_3 b} d^{b a_2 a_4} \right).
\eear
We introduce the notation
\bear
\Tgl &\;\equiv\;& 
\ln\left({ m^2-\hat t \over \sqrt{m^2 \hat s }}\right)\,
\; - \; {1 - i\pi \over 2}\; ,
\nn \\ \nn
\Ugl &\;\equiv\;& 
\ln\left({ m^2-\hat u \over \sqrt{m^2 \hat s }}\right)\,
\; - \; {1 - i\pi \over 2}\; ,
\nn \\ 
\Sgl &\equiv & -\frac{L_{\beta} + 1}{2}\;,
\eear
where the Mandelstam variables are given by
$$
\hat s = (p_1 + p_2)^2, \qquad \hat t = (p_1 - p_3)^2, \qquad 
\hat u = (p_1 - p_4)^2,
$$
and $\;L_{\beta} = \frac{1}{\beta}(1 - 2m^2 / \hat s)
\left( \ln{1-\beta \over 1+\beta} + i\pi \right)$. 
We also define 
\mbox{$\;\BA  \!\equiv \! \Tgl +\Ugl\,$,} \mbox{$\;\C  \!\equiv \! \Tgl - \Ugl\,$.}

Using the results for the one-loop eikonal integrals
from~\cite{KS} we calculate the matrices 
$\Gamma^{ij \tosv \gl\gl}$~\cite{inpreparation}.
For the $q\qbar$ channel we obtain
\be
\Gamma ^{q\bar q\tosv\gl\gl}\; =\;
{
%\small
{\alpha_s \over \pi} \,\left[\,
\left(
\begin{array}{ccc}
\rule{0em}{3.5ex} \;6\A\; & 0 & -\C \\
\rule{0em}{3.5ex} 0 &\, 3\A + {3\over 2}\BA \,& -{3\over 2}\C \\
\rule{0em}{3.5ex} -2\C & -{5\over 6}\C & \, 3\A + {3\over 2}\BA \,\\
\end{array}
\right) \, - \, \frac{4}{3}i\pi\,{\mathbf{\hat{I}}} \;\, \right]. 
}
\label{eq:qqbargamma}
\ee 
The matrix for the $gg$ channel has
the block form
\be
\Gamma ^{gg\tosv \gl\gl} \;=\;
{\alpha_s \over \pi} \; \left[\,
\left(
\begin{array}{cc}
\rule{0ex}{2ex}\; \Gamma_5\; & {\mathbf{\hat 0}} \\
\rule{0ex}{2ex}{\mathbf{\hat 0}}  & \;\Gamma_3\;\\
\end{array}
\right)
\, - \, 3i\pi\,{\mathbf{\hat{I}}}\;
\right]\, ,
\label{eq:gggamma}
\ee
where the five-dimensional matrix reads
\be
\Gamma_5 \; = \; 
{
%\small
\left(
\begin{array}{ccccc}
\rule{0em}{3.5ex}\;\;6\A\;\;   & 0     & 6\C    & 0   &  0  \\
\rule{0em}{3.5ex}0 & \;3\A + {3\over 2}\BA\; &    {3\over 2}\C  & 3\C   &  0  \\
\rule{0em}{3.5ex}\;{3\over 4}\C\;  & {3\over 2}\C     & \;3\A + {3\over2}\BA \;&
 0  & \;{9\over 4}\C\;   \\
\rule{0em}{3.5ex}0  & {6\over 5}\C     & 0     &\;\; 3\BA \;\;& {9\over 5}\C   \\
\rule{0em}{3.5ex}0  & 0     & {2\over 3}\C     & \;{4\over 3}\C   & 4\BA-2\A\; \\
\end{array}
\right)}\;
\ee
and the three-dimensional matrix $\Gamma_3$ is diagonal, 
\be
\Gamma_3 = 
%(\alpha_s / \pi) \; 
\mathrm{diag}\,(\,3(\Sgl+\Ugl)\,,
                \,3(\Sgl+\Tgl)\,,
                \,3(\Tgl+\Ugl)\;).
\ee
As in~\cite{KS}, the matrices $\Gamma$ contain terms from the one-loop
eikonal integrals for $2\to2$
scattering with massive particles in the final state and are shifted by half of the soft anomalous dimension for the Drell-Yan
cross section~\footnote{
Gauge
dependence in the one-loop integrals cancels against gauge dependence
of the incoming-jet factors (which corresponds to gauge dependence of the 
soft function in the resummed Drell-Yan cross section~\cite{KOScol, DMS}), 
leaving the results presented here gauge-invariant.}. 
The matrices~(\ref{eq:qqbargamma}),~(\ref{eq:gggamma}) complement the 
results of \cite{KOScol} for the case of pair-production of massive 
colour-octet particles with equal masses.

At the threshold, \mbox{$\shat \to {4m^2}$}, 
the soft anomalous dimension matrices approach the diagonal form,
\bear
\Gamma ^{q\bar q\tosv\gl\gl} &\to& 
{\alpha_s \over \pi}\, \mathrm{diag}\,(\gamma^g _1,\, \gamma^g
_2,\,\ldots,\, \gamma^g _8)\,, \nn \\
\Gamma ^{q\bar q\tosv\sq\sq} &\to&
{\alpha_s \over \pi}\,\mathrm{diag}\,(\gamma^q _1,\, \gamma^q
_2,\,\gamma^q _3).
\eear
The off-diagonal terms, proportional to $\C$, vanish like $\beta$ for $\beta \to 0$ and may be neglected.
Using  
$D^{(1)}_{gg\tosv \gl\gl,\, I} = 2\mre(\gamma^g _I)$
and $D^{(1)}_{q\bar q\tosv \gl\gl,\, I} = 2\mre(\gamma^q _I)$~\cite{KOSthr,KOScol}
we obtain 
\bear
\{D_{gg\tosv\gl\gl,\,I} ^{(1)}\} &=& \{0,-3,-3,-6,-8;-3,-3,-6\} \nn \\ 
\{D_{q\bar q \tosv \gl\gl,\,I} ^{(1)}\}&=& \{0,-3,-3\}.
\eear
Note that the values of the  $D^{(1)}$-coefficients are the negative
values of the quadratic Casimir operators for the SU(3) 
representations for the outgoing state. 
This agrees with the physical picture of the soft gluon 
radiation from the total colour charge of the heavy-particle pair 
produced at threshold~\cite{BCMNtop}.

\section{Numerical results}

In the phenomenological analysis we consider a wide range of 
gluino and squark masses. Left- and right-handed squarks of 
all flavours are assumed to be mass degenerate.
For the $\gl\gl$ production we vary~\footnote{For the highest
masses considered here, the experimental exploration will require
the integrated luminosity of $\ord(100 \fba^{-1}$).}
 the gluino mass, $\mgl$, 
between 200~GeV and 2~TeV. Similarly, for the $\sq\sqb$ production we take
200~GeV$\,<\msq <\,$2~TeV.
We present the results for
a fixed ratio of gluino and squark masses, $r=\frac{\mgl}{\msq}$,
and choose the following values $r=0.5, \; 0.8\; 1.2\; 1.6,\;2.0$.
The $\sq \sqb$ cross section accounts for production of all $\sq \sqb$
flavour combinations apart from the ones with scalar top particles.

The main phenomenological results of this letter are the
resummation-improved predictions for the  $pp \to \sq \sqb$ and $pp
\to \gl\gl$ total cross sections at  $\sqrt{S} = 14 \TeV$. The resummation-improved cross sections are 
obtained through matching the NLL resummed expressions with 
the full NLO cross sections
\bear
\label{hires}
\si^{\rm (match)}_{h_1 h_2 \tosv kl}(\rho, m^2, \{\mu^2\}) &\;=\;& 
\sum_{i,j=q,\qbar,g}\,
\int_{C_{\mathrm MP}-i\infty}^{C_{\mathrm MP}+i\infty}\;
\frac{dN}{2\pi i} \; 
\rho^{-N} \; f^{(N+1)} _{i/h{_1}} (\mu^2_F) \; 
f^{(N+1)}_{j/h_{2}} (\mu^2_F) \nn \\ 
\rule{0em}{2em}
&\times& \left[\; \left.
\sih^{\rm (res)}_{ij\tosv kl,N} (m^2, \{\mu^2\})
\; - \; \sih^{\rm (res)}_{ij\tosv kl,N} (m^2, \{\mu^2\}) \,
\right|_{\scriptscriptstyle({\NLO})}\; \right] \nn \\
&+& 
\rule{0em}{2em}\si^{\rm (NLO)}_{h_1 h_2 \tosv kl}(\rho, m^2, \{\mu^2\}) \,, 
\eear
where $\{\mu^2\}=\{\mu_F^2, \mu_R^2\}$, $\sih^{\rm (res)}_{ij\tosv
  kl,N}$ is given in Eq.~(\ref{eq:res:fact}) and  $ \sih^{\rm
  (res)}_{ij\tosv kl,N} \left. \right|_{\scriptscriptstyle({\NLO})}$ represents its
perturbative expansion truncated at NLO. The moments of the parton 
distribution functions $f_{i/h}(x, \mu^2_F)$ are 
defined in the standard way 
$$
f^{(N)}_{i/h} (\mu^2_F) \; \equiv \; 
\int_0^1 dx \, x^{N-1} \, f_{i/h}(x, \mu^2_F). 
$$
The inverse Mellin transform (\ref{hires}) is evaluated numerically using 
a contour in the complex-$N$ space according to the ``Minimal Prescription'' 
method developed in Ref.~\cite{Catani:1996yz}.

\begin{figure}[t]
\begin{center}
\epsfig{file=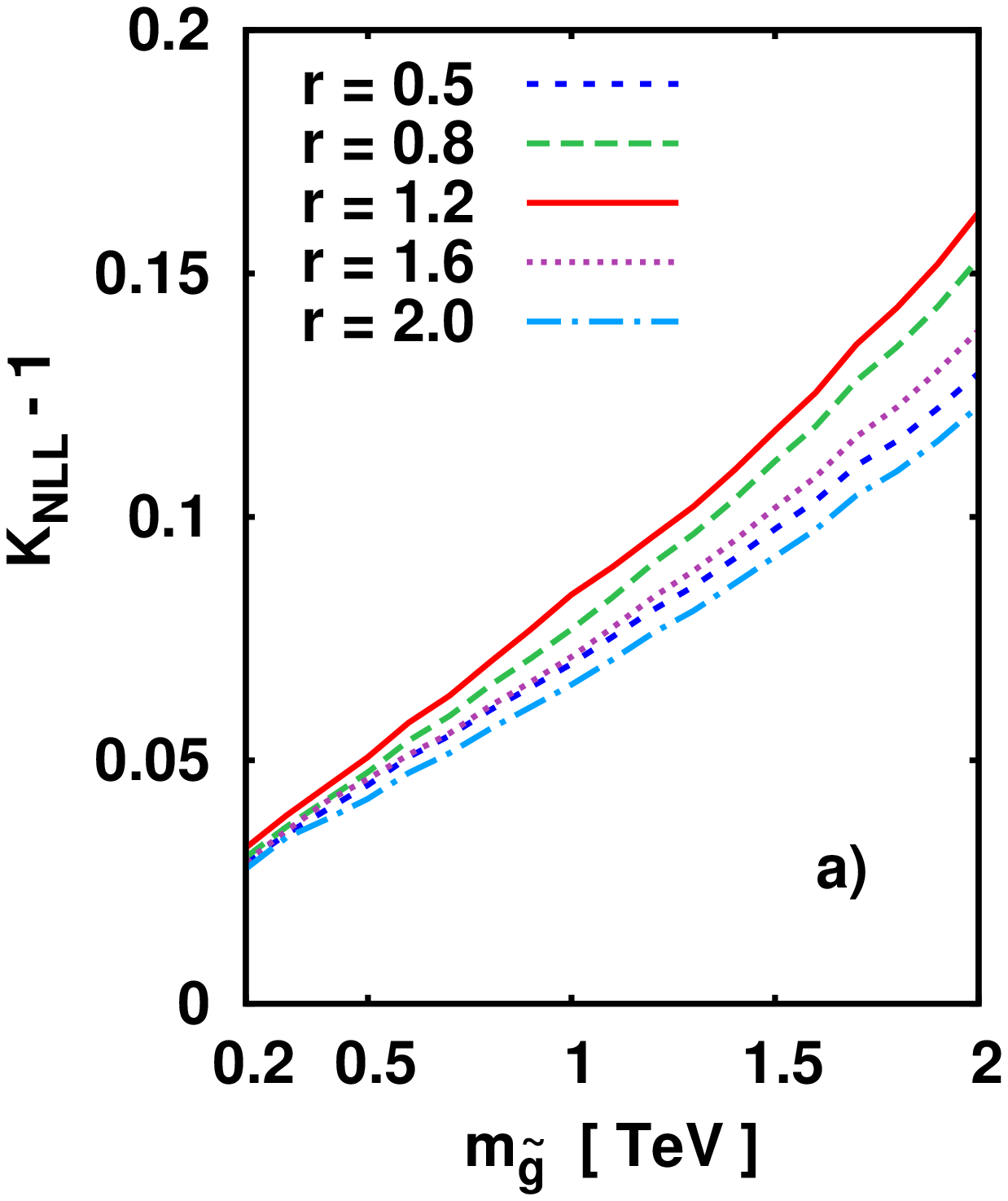,width=0.48\columnwidth}
\epsfig{file=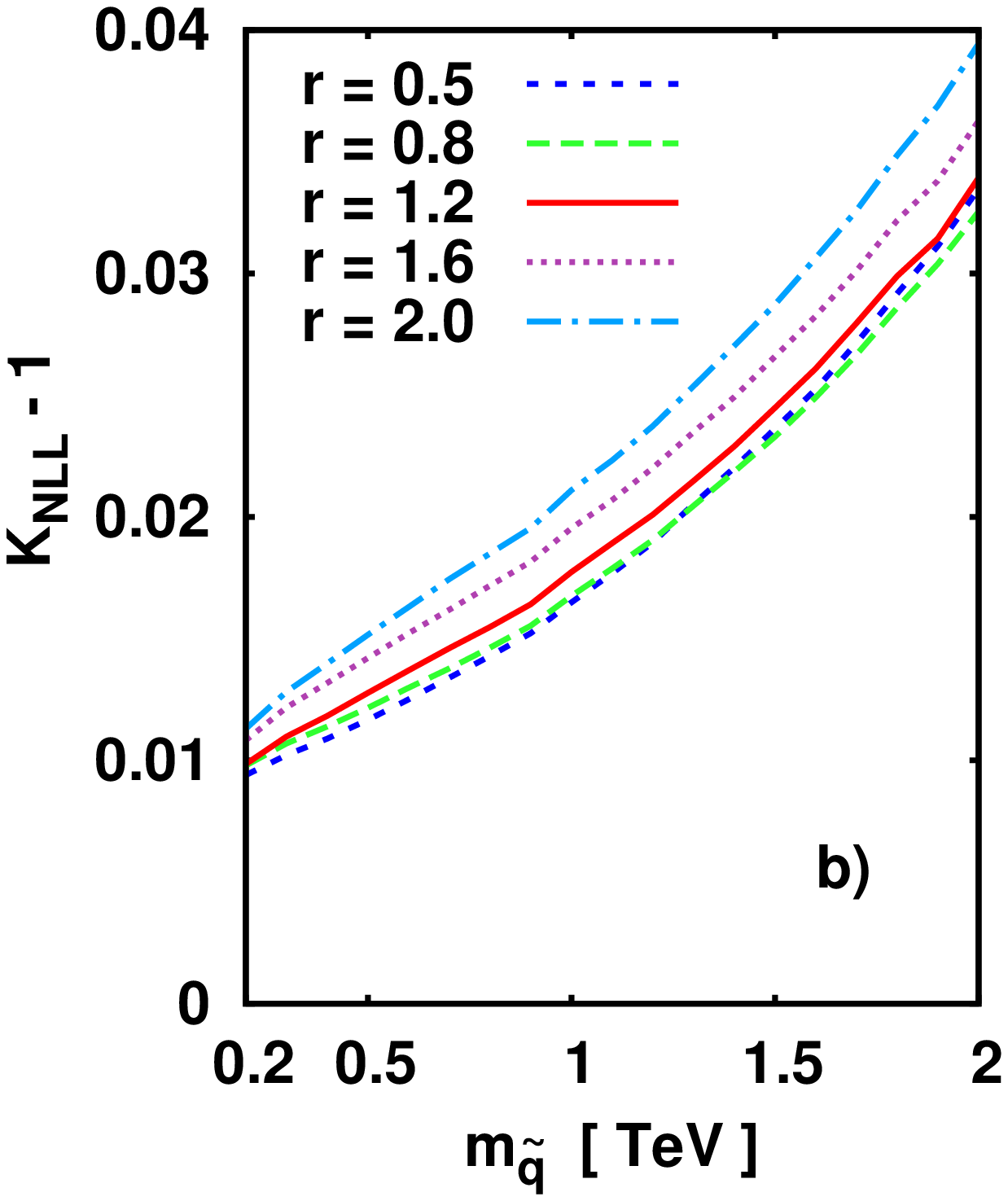,width=0.48\columnwidth}
\end{center}
\caption{\it 
The relative NLL correction
\mbox{$K_{\NLL}-1$}
for the $\gl\gl$ (a) and the $\sq\sqb$ (b) total production cross section 
at the LHC as a function of gluino and squark mass, respectively;
$r=\mgl / \msq$.}
\end{figure}

\begin{figure}[t]
\begin{center}
\epsfig{file=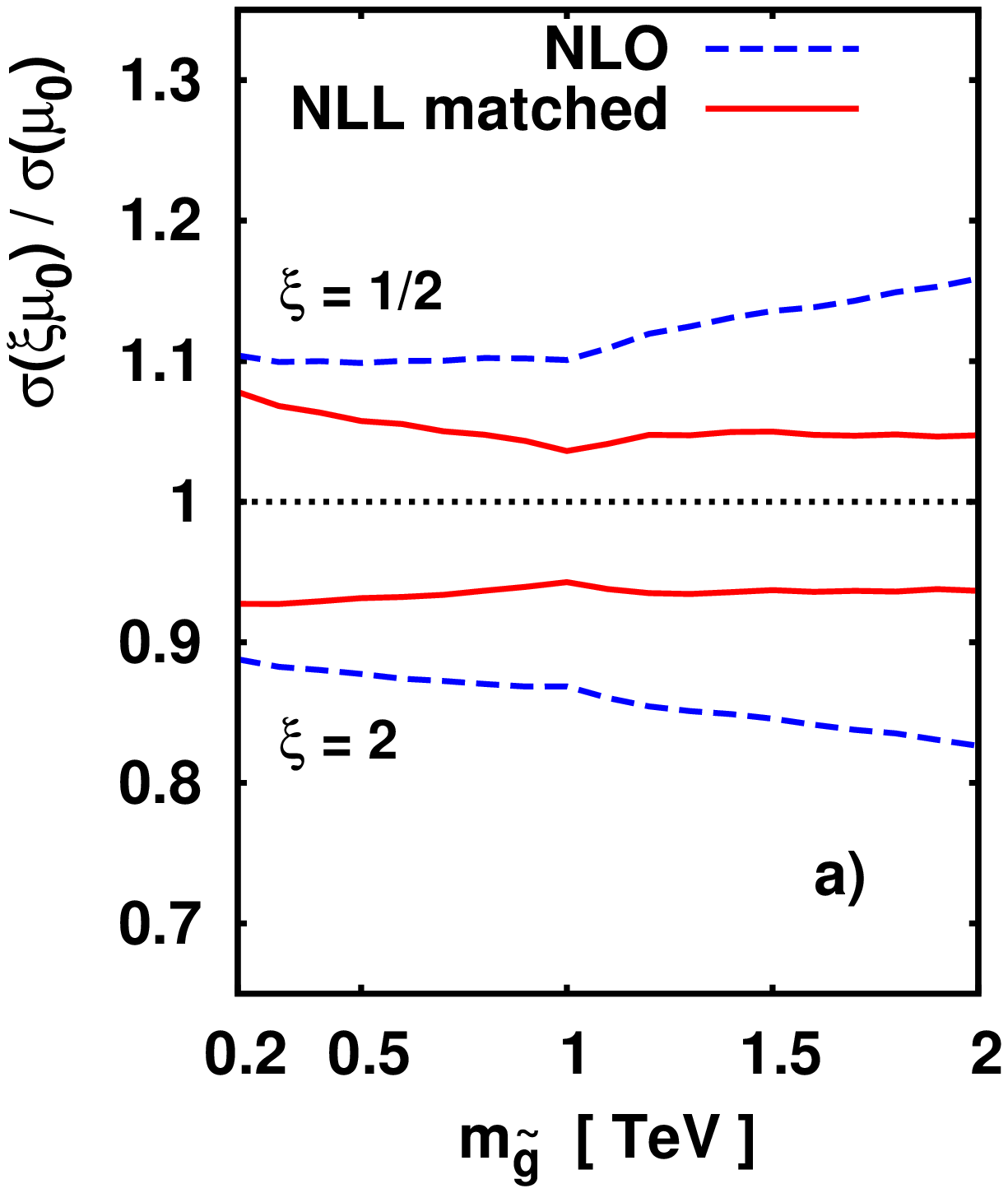,width=0.48\columnwidth} 
\epsfig{file=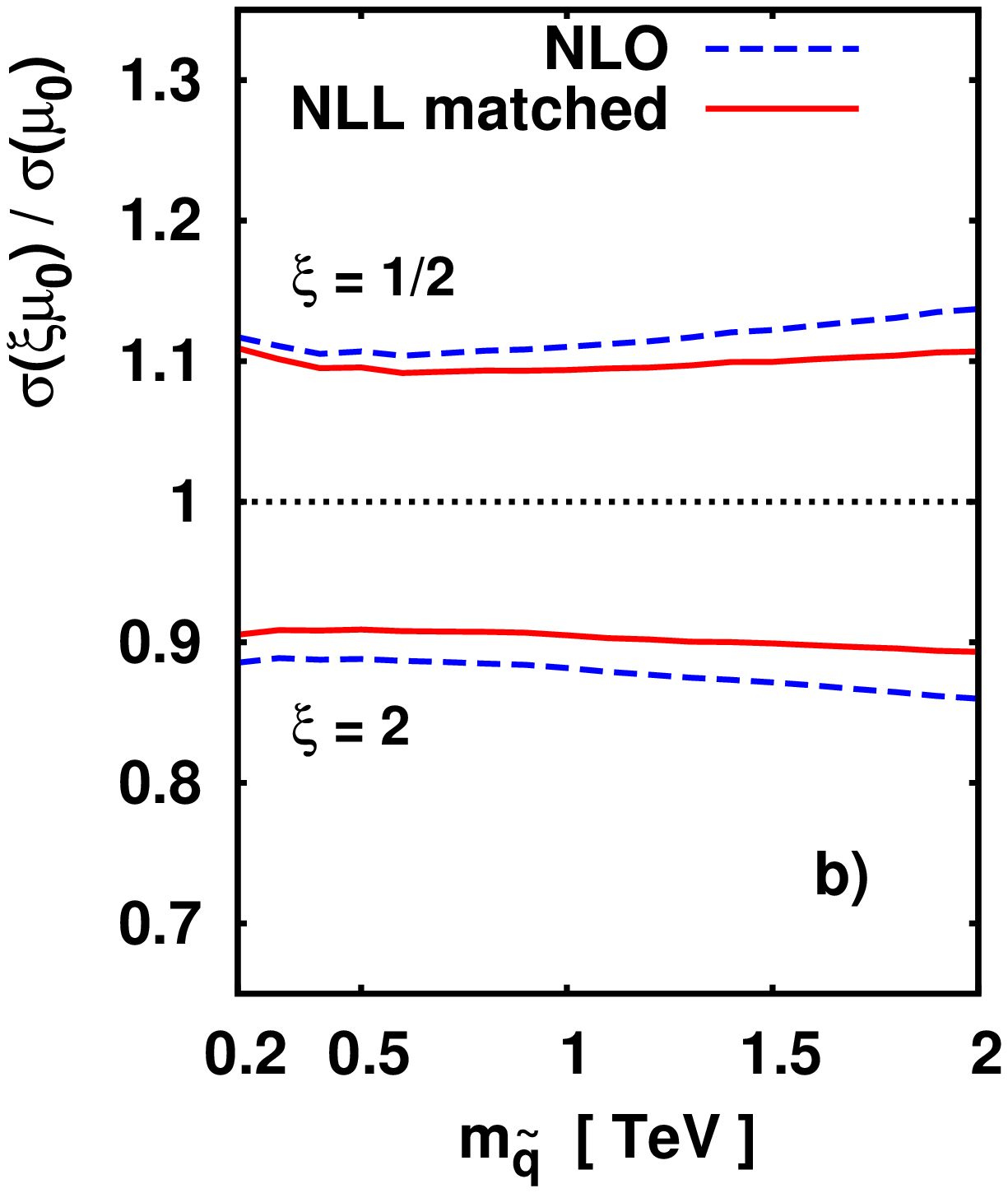,width=0.48\columnwidth}
\end{center}
\caption{\it Scale dependence of the total $\gl\gl$ (a)  
and $\sq\sqb$ (b) production cross section at the LHC 
(see the text for explanation).}
\end{figure}

The NLO cross sections are evaluated using \mbox{\sc
  Prospino}~\cite{prospino}, 
the numerical package based on
calculations employing the $\overline{\mathrm{MS}}$ renormalisation
and factorisation schemes.  We use the CTEQ6M~\cite{cteq6m}
parameterization of parton distribution
functions (pdfs) for all numerical predictions. Similarly to other
available pdfs, the assumption of five massless 
quark flavours active at large scales is made in the 
CTEQ6M parameterization. Consequently, in the NLO and NLL calculations we use 
the two-loop $\overline{\mathrm{MS}}$ QCD running coupling constant $\as$
with $\nf=5$ and $\Lambda^{(5)} _{\mathrm{QCD}}=0.226$~GeV. 
The effects due to virtual top quarks and virtual sparticles
in the running of $\as$ and in the evolution of pdfs are thus not included
in our predictions.
However, the value of the top mass, $m_t=175 \GeV$, enters the matched
NLL cross sections through the NLO corrections.

All numerical calculations were performed using two independent computer codes.
As a first check we reproduced numerically, and analytically, the LO
results~\cite{BHSZ2,prospino}. We also verified that the NLL resummed
expression in $N$-space generates the same 
logarithmic terms as these present in the Mellin transforms of the NLO
correction in the threshold limit given in~\cite{BHSZ2}.

The relative corrections from soft gluon resummation to the $\gl\gl$ 
and $\sq\sqb\,$ NLO production cross sections at the LHC,
$$
K_{\NLL}-1 \; \equiv \; \si^{(\mathrm{match})}/\si^{(\NLO)}-1\,,
$$
are presented in Fig.~1a and Fig.~1b, respectively. 
In the plots we set the scales $\mu_F=\mu_R=\mu_0$, where $\mu_0 = m_{\gl}$
($\mu_0 = m_{\sq}$) for the $\gl\gl$ production (the $\sq \sqb$ production).
$K_{\NLL}$ grows with the final-state mass and depends on the
mass ratio $r$ in a moderate way. 
The correction,
$K_{\NLL}-1$, reaches 16\% (8\%) for the $\gl\gl$~production with
$r=1.2$ and  
$\mgl= 2$~TeV (1~TeV), 
and 4\% (2\%) for the $\sq\sqb$~production with $r=2$ and 
$\msq= 2$~TeV (1~TeV). The stronger effect found in the $\gl\gl$ 
production follows from the dominance of the $gg\to \gl\gl$ channel, 
and hence more intense soft gluon radiation.

We also investigate the dependence of the matched NLL cross section
on the values of factorisation and renormalisation scales, in
comparison to the NLO cross section. To illustrate our results we
choose \mbox{$\mu=\mu_F=\mu_R$} and $r=1.2$. In Fig.~2a and  Fig.~2b we plot the ratios 
\mbox{$\sigma^{\NLO}(\mu=\xi \mu_0) /
\sigma^{\NLO}(\mu=\mu_0)$} and 
\mbox{$\sigma^{\mathrm{(match)}}(\mu=\xi \mu_0)
/ \sigma^{\mathrm{(match)}}(\mu=\mu_0)$}, obtained by varying
$\xi$ between $\xi=1/2$ and $\xi=2$, for $\gl\gl$ 
($\mu_0 = \mgl$) and $\sq\sqb$ ($\mu_0 = \msq$) 
production, respectively. Due to resummation, the scale sensitivity
of the $\gl\gl$ production cross section reduces significantly, by a
factor of $\sim 3$ ($\sim 2$) at $\mgl = 2$~TeV ($\mgl =
1$~TeV). At  $\mgl>1$~TeV the theoretical error of the matched NLL $\gl\gl$ 
cross section, defined by changing the scale \mbox{$\mu=\mu_F=\mu_R$} 
around $\mu_0 = m_{\gl}$ by a factor of~2, is around 5\%. 
In the case of the $\sq \sqb$ production,
the reduction of the scale dependence 
due to including soft gluon corrections in 
the theoretical predictions is moderate.

{
%\small 
\paragraph{Acknowledgments.} 
 We thank J.~Bartels and
G.~Sterman for comments on the manuscript. 
L.M.\ is supported by the DFG grant no. SFB 676.}

\end{document}